\documentclass[preprint,12pt]{aastex}
\usepackage{natbib}
\begin{document}
\newcommand{\be}{\begin{equation}}
\newcommand{\bel}[1]{\begin{equation}\label{eq:#1}}
\newcommand{\ee}{\end{equation}}
\newcommand{\bd}{\begin{displaymath}} 
\newcommand{\ed}{\end{displaymath}}   
\newcommand{\bea}{\begin{eqnarray}}
\newcommand{\beal}[1]{\begin{eqnarray}\label{eq:#1}}
\newcommand{\eea}{\end{eqnarray}}
\newcommand{\e}[1]{\label{eq:#1}}
\newcommand{\eqref}[1]{\ref{eq:#1}}

\newcommand{\bfr}{{\bf r}}
\newcommand{\bfrp}{{\bf r'}}
\newcommand{\Scal}{{\cal S}}
\newcommand{\Jcal}{{\cal J}}
\newcommand{\Bcal}{{\cal B}}
\newcommand{\Jbar}{{\bar J}}
\newcommand{\Sbar}{{\bar S}}
\newcommand{\Jcalbar}{{\bar \Jcal}}
\newcommand{\Ibg}{{I_{\rm bg}}}
\newcommand{\etal}{et~al.~}

\title{Dark Cloud Cores and Gravitational Decoupling from Turbulent Flows}
\author{Eric Keto\altaffilmark{1} and
George Field\altaffilmark{1}, 
}
\altaffiltext{1}{Harvard-Smithsonian Center for Astrophysics, 60 Garden Street, Cambridge
MA 02138}

\def\etal {et~al.~}
\def\diaz {N$_2$H$^+$}
\def\msun {$\rm M_\odot$}
\begin{abstract}

We test the hypothesis that the starless cores may be 
gravitationally bound clouds supported largely by thermal pressure by comparing
observed molecular line spectra to theoretical spectra produced by a
simulation that includes hydrodynamics, radiative cooling, variable
molecular abundance, and radiative transfer in a simple one-dimensional
model.  
The results suggest that the starless cores can be divided into two categories:
stable starless cores that are in approximate equilibrium and 
will not evolve to form protostars, and unstable
pre-stellar cores that are proceeding toward gravitational collapse and the
formation of protostars. 
The starless cores might be formed from the interstellar
medium as objects at the lower end of the inertial cascade of interstellar
turbulence.
Additionally, we identify a thermal instability in the starless cores. Under particular
conditions of density and mass, a core may be unstable to expansion if
the density is just above the critical density for the collisional coupling
of the gas and dust so that as the core expands the gas-dust coupling that cools
the gas is reduced and the gas warms, further driving the expansion.

\end{abstract}


\section{Introduction}

Observations of molecular clouds
show a power law dependence between the size (length and mass) scales of the clouds 
and the velocity dispersion within the clouds 
that extends from
the largest giant molecular clouds
down to the scale at which the turbulent velocities
become subsonic 
\citep{Larson1981,LKM1982,Myers1983,
SSS1985,Dame1986,FullerMyers1992}. 
The spectrum
matches that expected in a
turbulent cascade suggesting that
the clouds are hierarchical structures in a
supersonic turbulent flow
\citep{MacLowKlessen2004,
ElmegreenScalo2004,ScaloElmegreen2004}.
The power law relationships between the mass, length, and velocity dispersion also fit
the relation of virial equilibrium, $\sigma_v^2 \sim 2GM/L$ as if
the clouds were
gravitationally bound structures.
In a picture of the interstellar medium dominated by turbulence,
the relationship of apparent virial equilibrium
reflects a dynamic equipartition as a result of the coupling 
between kinetic and potential energies in
the turbulent flow rather than a static equilibrium within gravitationally bound clouds
\citep{Larson1981, BPVSScalo1999, Klessen2005}. 
While gravity may be largely responsible for generating the supersonic flows that form the clouds, 
the clouds themselves are the result of compression due to inertial forces in
the turbulent hydrodynamics.
Clouds are formed where converging streams in the turbulence 
create zones of high density gas, but these clouds formed by compression may just as
easily dissipate as the flows change direction and velocity on a crossing time scale.
In the inertial range of the interstellar turbulence, there are no 
clouds with an internal equilibrium between self-gravity and the
supporting forces of thermal, turbulent, or magnetic pressure.
Gravitationally bound clouds may form if
the compression is strong enough to boost the local gas
density beyond gravitational instability.
However, these 
clouds will be
short-lived, collapsing to form stars within a free fall time. 
In the theory of the turbulent interstellar medium, these are the 
only gravitationally bound clouds, and
star formation takes place within these collapsing clouds 
within one crossing time \citep{Elmegreen2000}. The inefficiency or slow rate
of star formation is then attributed to the inefficiency in the formation of these
gravitationally bound clouds in the turbulent interstellar medium.

While the theory of the turbulent ISM does not predict any molecular clouds
in equilibrium, the observed properties of the small molecular clouds
at the low end of the cloud mass spectrum known
as starless cores are remarkably well described as
discrete dynamical units of stable, bound gas with an internal balance
of forces. 
These starless cores 
are dense regions ($n_{\rm H_2} \sim 10^4$ to $10^6$ cm$^{-3}$) in dark clouds
with linear scales of tenths of pc, and total masses of a few solar masses. The
starless cores contain no infrared sources above the sensitivity level of
the IRAS satellite (about 0.1 L$_\odot$ at the distance of Taurus) and thus are
thought to be sites of possible future rather than current star formation 
\citep{MyersLinkeBenson1983, MyersBenson1983, BensonMyers1989, Beichman1986, Ward-Thompson1994, Tafalla1998, LeeMyers1999, LeeMyersTafalla2001}. 

Observations of dust column density in the starless cores appear
approximately as
expected for cores in hydrostatic equilibrium. The observed density profiles are
characterized by a core-envelope
structure with an inner region of weakly
decreasing density and a surrounding envelope with a steeper density
gradient
\citep{Ward-ThompsonMotteAndre1999, Bacmann2000, Shirley2000, Evans2001, ShirleyEvansRawlings2002, Young2003, Keto2004}. 
Detailed observations of some individual dark clouds show density
profiles that appear to match within a few percent those 
of pressure-confined, hydrostatic spheres 
(Bonnor-Ebert spheres)\citep{Bonnor1956,AlvesLadaLada2001,Tafalla2004}. 
While the observed morphologies of the starless cores are
not always spherical, and aspect ratios of two to one are common,
the observed density profiles do not deviate significantly from
those expected in near-equilibrium.

Radio frequency spectral
line observations that reveal the gas velocities within the cores confirm this 
near-equilibrium state.
The observed spectral lines of all the starless cores
are quite narrow with no wings
and nearly thermal widths, an indication
that any gas velocities are a few
tenths the sound speed or less
\citep{Zhou1994, Wang1995, Gregersen1997, Launhardt1998, GregersenEvans2000, LeeMyersTafalla1999, LeeMyersTafalla2001, AlvesLadaLada2001, LeeMyersPlume2004, Keto2004}.
The cores contain a variety of gas motions,
for example contraction, expansion, or
both within different regions of the core, but
the observation that the gas velocities are all subsonic implies that the inertial forces are
small, and that there is 
near balance between the thermal
and gravitational forces.

A state of near equilibium as indicated by both the dust and spectral line
observations
would allow the starless cores to be long-lived, apparently
at odds with the transience required of the clouds in the theory of the
turbulent interstellar medium.
However,
there might be no conflict between the two concepts of clouds as either transient 
entities or
as pressure supported dynamical units
if the larger clouds are within and the starless cores are below the 
inertial range of the supersonic turbulence.
The idea that the
starless cores with their narrow line widths are clouds at the subsonic bottom of
the turbulent cascade has been discussed in the literature since
Larson's description of the size-line width relationship
\citep{Larson1981, Padoan1995, Goodman1998, VSBPKlessen2003}. 
What has evolved since the earliest description of the interstellar medium
as a turbulent cascade is the concept of the larger scale
clouds in the inertial range of supersonic turbulence as purely transient
phenomena entirely out of equilibrium.  The question now is whether all clouds
in the interstellar medium including the starless cores are transitory.

We propose to test the hypothesis that the starless cores
are clouds in quasi-equilibrium by comparing against
observations the predicted characteristics of model near-equilibrium cores. 
If the modeled cores provide an adequate
description of the observations, then we would conclude that quasi-equilibrium
structures can and do exist in the interstellar medium. Given that the
interstellar medium is dominated on larger scales by the structures
of supersonic turbulence, then the starless cores must be 
the members at the bottom of the hierarchy, with scales below the sonic scale.

A similar approach of comparing model clouds in the turbulent hierarchy with
observations was discussed in \citep{BPKlessenVS2003}. That research examined the density
structures of a number of clouds produced in a numerical simulation of
interstellar supersonic turbulence and compared their structures to the
density profiles of pressure-supported, self-gravitating clouds. While none of
the model clouds was in equilibrium, a number of them had density structures
that matched those expected of clouds in equilibrium. Thus one would conclude
that observations of the density structure of clouds is an ambiguous
test of their state of equilibrium. While that comparison was motivated by
observations of the dust continuum in starless 
cores, an observation that
typically provides only the density structure, there are many observations of
molecular lines in starless cores that provide information on the velocity structure, the
temperature, and the chemistry of the cores. With this additional
information, the comparison is less ambiguous. For example, the clouds in
figures 9 of \cite{BPKlessenVS2003} would never be confused with equilibrium
structures because of the supersonic velocities within the model
clouds. The dynamics of clouds with supersonic velocities are
dominated by inertial forces and not their
internal pressure. 
Therefore "observations" of these model clouds 
that included spectroscopy would indicate that they cannot 
be in equilibrium.

Our research thus aims to reduce the ambiguity of the comparison
by considering information that could be derived from molecular line 
observations of the starless cores in addition to the density structure derived
from observations of the dust continuum.
Since we model only individual clouds in quasi-equilibrium rather
than the entire larger scale turbulent cascade, we can include in our models physics
such as the radiative equilibrium of the dust and gas, and the depletion of molecules from
the gas phase by freezing onto dust grains, that with the present computing power,
could not be fully incorporated into
a three-dimensional hydrodynamic simulation.
Finally, our models can be run at much higher spatial resolution. The level of
detail provided by the additional physics and resolution improves the
predictive power of our models and reduces the ambiguity of the comparison
allowing us to improve on the results of \cite{BPKlessenVS2003}.

The results of our comparison suggest that the observations 
of the starless cores are well explained by models in
close equilibrium. For a given external pressure, the equilibrium may be 
stable or unstable to gravitational collapse
depending on the mass of the core. 
It seems reasonable to suppose that there will be a distribution 
in the initial masses of the cores because
the cores are formed at the subsonic scale at the bottom 
of the turbulent cascade that itself
follows a power law distribution in size scale in the inertial regime.
Cores that are more massive and dense 
at the time of formation ($n({\rm H}_2) > 10^5$ cm$^{-3}$
and a few M$_\odot$) will be 
ultimately unstable to gravitational collapse. Less dense examples 
may be in stable equilibrium, but stability does not imply that the cores
must be static.
For example, stable cores may oscillate 
in size and density around an equilibrium mean.
While the unstable cores will
follow a progression toward greater densities and infall velocities, 
there is no evolutionary sequence linking the stable to the unstable cores.
The less dense stable cores will not of themselves progress toward greater density 
and ultimate collapse, but may
persist indefinitely until external conditions change. 
In this picture there are two classes of cores and 
some of the inefficiency or slow rate of star formation will be
due to the formation of a significant fraction of the population of
cores that are stable and not inclined to
evolve to form stars.

\section{Physics of the Starless Cores}

\subsection{Hydrostatic and Radiative Equilibrium}

Given the power law relationship, $\sigma_v \sim r^{1/3}$, 
between the turbulent velocity, $\sigma_v$,
and the cloud
length scale, $r$,
the size scale where $\sigma_v$ is the sound speed defines the
fragmentation scale of the turbulent flow 
\citep{Larson1981, Padoan1995, Goodman1998, VSBPKlessen2003}. 
Below this
length scale, the turbulent velocities are subsonic, and the
turbulent energy is less than the thermal energy. The fact that the
thermal pressure dominates the dynamics
immediately leads to a qualitative
description of the structure of subsonic cores that
matches the observed properties of the starless cores.  
Subsonic turbulence can not produce substructure within the core, and therefore
the density profile must be smooth and monotonic. 
Moreover, the 
density gradients must be characteristic of thermal pressure support
because the inertial
forces are small, unless the cores are in free-fall collapse.
Thus there must be approximate force balance with
thermal pressure rather than turbulent pressure
as the dominant support against gravity.

If one imagines a boundary to the core,
this leads to a description of the
starless cores as
hydrostatically supported spheres,
confined by an external pressure. If the gas is isothermal these spheres
are generally referred to as
Bonnor-Ebert (BE) spheres 
described by the solution of the Lane-Emden equation, truncated at
some radius, with an external pressure that is set 
equal to the internal pressure,
$P = nkT$, at the truncation radius \citep{Bonnor1956}.
Because the cores are embedded in larger scale molecular clouds,
they will be continuous
with the surrounding molecular gas with no sharp boundaries
in pressure, temperature, or density. 
In particular, the notion of the starless cores as
BE spheres does not require or imply a hot, rarefied interstellar
medium around the cores. 
Because the velocity dispersion in the molecular clouds increases with
the length scale, there will be some scale around a core where
the turbulence becomes supersonic, thermal pressure no longer dominates,
and the dynamics are dominated by the inertial forces. This scale
defines the boundary of the core. However, this definition 
is more conceptual than precise, and in practice, the
boundaries in our models are chosen more simply, for example, to
set the total mass of the core.

Whether a core in equilibrium can be long-lived depends on its dynamical stability
against gravitational collapse. 
The gravitational stability of an isothermal sphere is described by Bonnor's criterion
that the pressure at the boundary of the core should increase as the volume of the core
is reduced. 
Similar to the case of an isothermal
sphere, the static stability of a non-isothermal
sphere, that might include some non-thermal internal pressure, can be determined by
the sign of the change in boundary pressure as the volume of the core is reduced.
Because the temperature variation in the starless cores varies by only several degrees 
around an average temperature of about 10 K, 
a stability analysis of the non-isothermal models shows 
a result that is similar to the isothermal case, although the critical 
values are somewhat different.

In the next section, we describe our static model for starless cores as 
spheres in hydrostatic and radiative equilibrium. This model serves as
the starting point for the calculations of the 
hydrodynamic evolution, but also provides a reasonable approximation to the
internal structure of a starless core at any point in its evolution.
This follows 
because the cooling time of a core is much shorter than its dynamical
time, and because
the cores are always in near equilibrium, unless in gravitational free-fall.
This will be justified later with the results of the evolutionary calculations.

\section{Models of Hydrostatic Spheres in Radiative Equilibrium}

The hydrostatic equilibrium is defined by the usual equations for the
force balance between pressure and gravity, the mass of the sphere, and
the equation of state \citep{Chandrasekhar1939}, 
\be
{{dP} \over {dr}} = {{GM(r)\rho}\over{r^2}} \e{hydrostatic}
\ee
\be
{{dM} \over {dr}} = 4\pi r^2\rho \e{mass}
\ee
\be
P = kT\rho/m \e{eos}
\ee
where $m$ is the average mass per molecule, including helium and heavier elements. 
These equations are non-dimensionalized
by defining,
\be
\theta = T/T_R \e{tscale}
\ee
\be
\sigma = \rho/\rho_c \e{rhoscale}
\ee
where $\rho_c$ and $T_R$ are a reference density and temperature. We use
the density at the center of the sphere
because the BE spheres are conventionally
characterized by the central density, but we use the temperature at the
boundary because the boundary temperature, unlike the central temperature,
is nearly independent of the mass and density of the sphere.
A linear scaling constant, $a$, related to the ratio of thermal
to gravitational energy, is used to non-dimensionalize the length and mass.
\be
a = \bigg({{kT_R}\over{Gm4\pi \rho_c}}\bigg)^{1/2} \e{ndratio}
\ee
\be
\xi = r/a \e{rscale}
\ee
\be
z = {{M(r)}\over{4\pi a^3\rho_c}} \e{mscale}
\ee
From the equations above, one derives two coupled ordinary differential
equations, which
we solve with a second-order Runge Kutta algorithm,
that define the density of a sphere in hydrostatic equilibrium
for the temperature profile, $\theta$ determined by radiative equilibrium.
\be
{{dz}\over{d\xi}} = \xi^2\sigma \e{ode1}
\ee
\be
{{d\sigma}\over{d\xi}} = -{{1}\over{\theta}}\bigg(z{{\sigma}\over{\xi^2}} +
\sigma{{d\theta}\over{d\xi}}\bigg ) \e{ode2}
\ee

Closely following several previous papers 
on the radiative equilibrium of the dark cloud cores,
we calculate the radiative equilibrium of the gas assuming that
the gas is heated by cosmic rays, cooled by molecular line radiation,
and either heated or cooled by collisional coupling with the dust
depending on whether the dust is hotter or cooler than the gas.
\citep{Larson1973, Larson1985, Evans2001, ShirleyEvansRawlings2002, Zucconi2001, StamatellosWhitworth2003, Goncalves2004}. 
For the dust,
\be
\Gamma_{ISR} - \Lambda_d = 0 \e{dusteq}
\ee
where, $\Gamma_{ISR}$ is the rate at which dust grains are heated by 
the external interstellar
radiation field, and $\Lambda_d$ is the rate at which grains cool by 
blackbody radiation modified by the dust emissivity. 
The equilibrium gas temperature
is defined similarly as the temperature at which the net gas cooling
rate ${L}$ is zero,
\be
{L} = \Gamma _{CR} - \Lambda _{line} - \Lambda _{gd} \e{gasequil}
\ee
where $\Gamma_{CR}$ is the rate of heating by cosmic rays 
(equation \ref{eq:crheating}),
$\Lambda_{line}$ is the rate of cooling by molecular line radiation
(equation \ref{eq:linecooling}),
and $\Lambda_{gd}$ is the rate of energy transfer between the
gas and the dust by collisions (equation \ref{eq:gdcoupling}). 
In equation \ref{eq:dusteq}, the lack 
of a corresponding term for the collisional coupling
is an approximation reflecting the much higher
rate of energy transfer by the dust through absorption and
emission of radiation  \citep{Goldsmith2001, Goncalves2004}.
This allows the dust temperature to be calculated independently
of the gas temperature. In equilibrium, 
the total radiative gas cooling rate, ${L} = 0$, but during the hydrodynamic
evolution where the radiative gas cooling rate enters into the equation for
the change in internal energy of the gas (equation \ref{eq:lenergy}), 
this rate 
is not generally zero because the gas temperature is usually different
from its value in radiative equilibrium because of compression
or expansion of the gas.

The dust temperature at each point in the cloud 
is calculated assuming radiative equilibrium
between the incoming interstellar radiation field that heats the dust
and the infrared emission that cools the dust
\citep{Zucconi2001, Goncalves2004}. 
Our calculation uses the parameterization of the interstellar radiation
field of \cite{Black1994} and the parameterization of the dust opacities of 
\cite{OssenkopfHenning1994} that are derived and described in \cite{Zucconi2001}. 
The incoming interstellar radiation is attenuated by the overlying molecular gas
and
the calculation of the incoming radiation field
involves the optical depths from each point in the core
to the surface of the core defined by the truncation radius for
the non-isothermal BE sphere, and averaged over rays at all angles. 
Because of the low temperatures in the cores, the dust radiates
primarily in the far infrared, and the core is assumed to be
transparent to this longer wavelength radiation.

Following \cite{FalgaronePuget1985} or \cite{Goldsmith2001}, we set
the rate of energy transfer from cosmic rays into the gas as,
\be
\Lambda_{cr} = 10^{-27} n_{H_2} {\rm ergs\ cm}^{-3}{\rm s}^{-1} \e{crheating}
\ee

The molecular line cooling follows 
\cite{Goldsmith2001} and uses the parameterized cooling functions for
standard abundances that are derived in that study. 
\be
\Lambda_{line} = \alpha(T_g/10K)^\beta {\rm\ ergs\ cm}^{-3}{\rm\ s}^{-1} \e{linecooling}
\ee
where the parameters $\alpha$ and $\beta$ are given in Table 2 of
\cite{Goldsmith2001}.

The parameterization
is based on the large velocity gradient approximation,
assuming a velocity gradient of 0.5 or 1.0 kms$^{-1}$ pc$^{-1}$. In
the model cores in our study, the velocity gradients are almost never
this high. However, the large number of molecules with different
abundances and transitions makes the parameterization insensitive
to the exact velocity gradient. Molecular line radiation escapes the
cloud, cooling the gas, through those transitions that have
optical depths of approximately unity. As the density increases
and the core becomes optically thick in a lower quantum transition
of a more abundant species such
$^{12}$CO(1-0), the radiation escapes through higher transitions of
$^{12}$CO and through transitions of less abundant species such as
$^{13}$CO. Thus while the optical depth of one transition itself is a sensitive
function of the assumed velocity gradient, the total cooling rate
is a less sensitive function since over a wide range of density
and velocity gradient there are always some transitions of optical 
depth unity. 

Energy is transferred between the gas and dust by collisional coupling.
Assuming a Maxwellian distribution of
random velocities in the gas \citep{BurkeHollenbach1983} and a
grain size distribution 
that we assume follows the -3.5 power of the grain radius from
0.03 $\mu$m to 0.3 $\mu$m 
\citep{KruegelSiebenmorgen1994}
and a gas to dust mass ratio of 100,
\be
\Lambda_{gd} = 10^{-33}n_{H_2}^2 T^{1/2} (T_g - T_d) {\rm\ ergs\ cm}^{-3}{\rm\ s}^{-1} \e{gdcoupling}
\ee
This is essentially
the same rate as used by
\cite{Zucconi2001},
\cite{Goldsmith2001}, and \cite{Goncalves2004}. 

Any two of the three
parameters, central density, total mass, or external pressure along
with a temperature profile is
sufficient to specify the solution of the Lane-Emden equation for hydrostatic
equilibrium. The temperature profile may then be determined for a given density
profile by solving for the radiative equilibrium of the gas and dust.
A few alternating solutions of the hydrostatic and radiative equilibrium equations
results in a non-isothermal sphere characterized by a central
density and total mass that is in equilibrium with an external pressure and
interstellar radiation field. 

\subsection{Results of the static modeling}

The stability of the non-isothermal cores is summarized in 
figure \ref{fig:pvplot_variable_T}, and compared to the
stability of an isothermal core in
figure 
\ref{fig:pvplot_constant_T}.
In each figure, the equilibria on the
curve to the left of the maximum equilibrium pressure
are unstable to gravitational collapse. 
Because the central density correlates with the external pressure and
inversely with the volume, the 
stability of a sphere can also be defined in terms of its central density.
The curves in figures \ref{fig:pvplot_variable_T}, \ref{fig:pvplot_constant_T},
can be thought of as curves of equilibrium
pressure and volume parameterized by density.
Some representative central 
densities are plotted along the curves.  A comparison of the curves shows that the 
critical boundary pressure or central density
at which the equilibrium becomes unstable is
different in the two cases, but the character of the curves is the same.

If the gas has additional internal energy such as might be
provided by small scale subsonic turbulence or magnetic fields, the critical
densities are higher. 
The stability diagram in figure 
\ref{fig:pvplot_nonthermal_pressure} 
shows the effect of additional internal energy equal to one-quarter of the thermal
energy of the gas. 
The additional
non-thermal energy improves the stability of spheres over their thermal
counterparts.

The gas and dust temperatures, and gas density for equilibrium spheres with
central densities of $10^4$ cm$^{-3}$ and $10^6$ cm$^{-3}$ are shown in figures
\ref{fig:tdv_stable} and \ref{fig:tdv_unstable}. 
These two
densities 
lie on either side of the critical density for gravitational
instability for a core with a total mass of 5 M$_\odot$.
The figures illustrate the differences in structure
predicted for stable and unstable cores. The most significant
differences in the two cases are in the degree of central condensation and in the
temperature.

The density profile of
the unstable cores has a smaller radial width or higher degree 
of central concentration than the
stable counterparts. 
In the observational literature, the width is usually 
defined as the radius of the central region of the
observed core inside of which the density is approximately constant and outside of which the density
falls as a power law. For example, the density profile may be described as
$n \sim 1/(1 + r^\alpha)$ \citep{Tafalla2004} or more simply in terms of two zones, an inner
zone of constant density within a radius $r_0$ and a surrounding power law envelope 
\citep{Ward-Thompson1994}.
For cores with total masses of several M$_\odot$,
the width of the constant density region is about 2500 AU in the case of
unstable pre-stellar cores, and about twice that or 5000 AU in the 
stable starless cores. 

The gas temperature 
in a stable sphere with lower density,
is approximately constant at 10 K. Although the dust
temperature decreases at smaller radii because the overlying cloud shields
the dust in the center from external starlight, the gas and dust are not
collisionally coupled, and the dust temperature has little effect on the
gas. The gas cools primarily by molecular line radiation, and the increasing
optical depth to line radiation causes a slight inward rise in temperature.

In the unstable core with high density, the increased optical depth to
the external starlight reduces the heating of the dust.
Because the dust is cooled by optically
thin far infrared radiation, the cooling rate remains constant, and the dust
becomes colder toward the center. 
At central densities $> 10^6$ cm$^{-3}$ the extinction through the core causes
the dust temperature to decrease from 17 K at the boundary to below 8 K at
the center. 

The unstable core has a high enough density that 
the gas becomes optically thick to molecular line radiation 
and in addition begins to collisionally couple to the dust.
Towards the center of the core, at densities  $> 10^5$ cm$^{-3}$, the
gas and dust are collisionally well coupled and the gas is cooled predominantly
through the dust. In the densest part of the unstable core where the coupling is most
efficient, the gas temperature
approaches the dust temperature, around 8 K. 
Outside of the center, the collisional coupling and hence the gas cooling is reduced 
as the density
decreases, and the gas temperature climbs to about 13 K. At still larger radii 
the optical depth to the surface of the core becomes low enough that
the gas can cool efficiently through molecular line radiation
and the temperature falls to just below 10 K. 
The critical density of the non-isothermal cores in radiative equilibrium
is above that
of the 10K isothermal core even though the central temperature is
lower in the non-isothermal core. This is because the 
cores in radiative equilibrium have a higher average
temperature.

It is coincidental to the condition of radiative equilibrium that the densities
at which the cores develop a strongly varying temperature profile are also
the densities at which the cores become unstable to gravitational collapse.
Nonetheless this circumstance results in a useful division of the population of
the starless cores into two classes.
In the first category are cores that are true
``starless'' cores in that they will not of their own contract to form
protostars. In the second category are
``pre-protostellar'' or ``pre-stellar'' cores
that have not yet formed a protostar, but are unstable to
collapse of the core and the formation of a protostar within a free-fall time.
We will combine the observational and theoretical descriptions and
refer to the first category as stable starless cores and the second
as unstable pre-stellar cores. In the absence of changing external conditions,
there is no evolutionary path linking the two
categories of cores. Stable cores do not evolve to become unstable.

\subsection{Comparison to observations}

Examples of well-studied stable starless cores include
B68 and L1517B. Well known cores in the unstable pre-stellar category include L1544 and L1521F. 
Estimates of the central density in the stable starless core L1517B are 
$2.2 \times 10^5$ from dust \citep{Tafalla2004} and $10^{5.1 \pm 0.3}$ 
from the molecular line \diaz\ \citep{Keto2004}. 
The width of the density profile
in L1517B is  35$^{\prime\prime}$ (0.025 pc) from dust \citep{Tafalla2004} and
$0.022 \pm^{0.008}_{0.016}$ pc from \diaz .
In the unstable pre-stellar core L1521F, 
the central density is estimated at $10^6$ cm$^{-3}$ with a width of 0.017 pc for the
central region of the core \citep{Crapsi2004}. The density structure of L1544 appears
similar with a
central density  $10^{5.7 \pm 0.4}$ cm$^{-3}$ and width 
$0.004 \pm 0.002$ pc from the observations of \citep{Keto2004} 
and  $10^6$ cm$^{-3}$ with width 20$^{\prime\prime}$ (0.015 pc) 
from the observations of \citep{Tafalla2002}.
In general, the stable, starless cores have lower densities with broader density profiles than
the unstable, pre-stellar cores.

The theoretically predicted decrease in dust temperature has been
observed in several cores.
Infrared observations show a dust temperature profile that
decreases from about 15 K at the edge of the core to 8 K in the center 
\citep{WardThompson2002, Pagani2003, Pagani2004}.

Molecular line observations of 
cores thought to be stable and not currently evolving toward star formation, 
such as L1517B and B68 show
nearly constant gas
temperatures with little variation across the core. Observations
of L1517B in NH$_3$ indicate a gas temperature of 10 K (L1517B, \cite{Tafalla2004})
while observations of \diaz\ indicate an average temperature of 
$9 \pm^{7}_{2}$ K \citep{Keto2004}.
Temperature estimates of B68 range from
10 to 16 K \citep{AlvesLadaLada2001, Lada2003}.  

Observations of the core L1544
indicate an average temperature of 8.75 K \citep{Tafalla2002} from 
NH$_3$ lines and $11 \pm 4$ K \citep{Keto2004} from \diaz , both
consistent within the observational uncertainties with the models,
but the spectral line 
observations lines do not have the angular resolution or the sensitivity to
define the temperature profile of the gas with the detail of the models. 
In general, temperature estimates of the stable, starless cores 
are a few degrees warmer than those of the unstable, pre-stellar cores,
consistent with the higher temperatures predicted by the models.

In order to understand the velocity structure of the cores and to interpret spectral
line observations it is necessary to model the hydrodynamic evolution of the cores.
In the sections below we develop the hydrodynamic model and present some examples
of the evolution of cores. We will find that 
both classes of cores, stable starless and unstable pre-stellar,
may have coherent internal velocities. The stable
starless cores can exhibit a variety of oscillatory motions, expansion and contraction,
that do not result in the collapse or dissipation of the core. The unstable pre-stellar
cores have inward velocities that will ultimately evolve to gravitational
collapse at free-fall velocities.
The two types of cores can be distinguished by their
spectral line profiles and widths. 
The spectral lines of unstable pre-stellar cores tend to show
strongly asymmetric profiles split by self-absorption. 
Also, the spectral linewidths of volatile molecular species
are broader in the unstable pre-stellar cores than they are in the stable starless
cores. 

\section{Hydrodynamic Model}

To model the evolution of the starless cores
we use a simple one-dimensional hydrodynamic code with a first order
Lagrangian discretization and Richtmyer-Von Neumann pseudo-viscosity.
While this method is not as sophisticated as the
piece-wise parabolic method (PPM) \citep{ColellaWoodward1984}
used in one of the previous
dynamical studies of BE spheres \citep{FosterChevalier1993},  the
improvement in computer speed allows our hydrodynamic code to
be run with a factor of 10 more grid points than in that previous
study, and this allows better spatial resolution despite the
inherent smoothing in the 
pseudo-viscosity method. Also, in our study the resolution is not as
important as in the Foster and Chevalier study.
First our investigation concerns only the stability of starless cores.
Thus we follow the evolution of cores only to a scale of AU which
is sufficient to determine the stability and evolutionary path
of the core. 
For the typical conditions in starless cores, most of the velocities are
subsonic on this scale.
In contrast the previous calculations of 
Foster and Chevalier also sought to characterize
the asymptotic behavior of the flow at the time of 
formation of a point source at the flow center.
Near a gravitational point source the flow can
accelerate to arbitarily high Mach numbers. In contrast to determine
the fate of a core, our simulations 
need not follow evolution much past the point when the flow has accelerated to more 
than unity Mach number. Supersonic infall velocities imply that
the core is in free-fall collapse, and its fate is sealed.

The typical core size in our models is 0.2 pc. The grid is initially
linear with approximately 5000 points yielding a resolution of about 10 AU.
The sound speed is about 0.2 km$^{-1}$ so the Courant condition implies
a maximum time step of about 250 yrs. The simulations are run with
a time step of 1/10 the maximum indicated by the Courant condition,
initially about 25 yrs. At the end of the evolution of an unstable core,
the typical maximum spatial resolution in the center of the core is less than one AU 
with a time resolution of less than one year.  As will be discussed below, the
limiting resolution in these simulations is not that of the hydrodynamics,
but of the radiative transfer.

The hydrodynamical equations in Lagrangian form are,
\be
r(s,t) = s + \int_0^t v(s,t) dt \e{lposition}
\ee
\be
{{1}\over{\rho(r,t)}} = {{1}\over{\rho(s,0)}}
\bigg({{r(s,t)}\over{s}}\bigg)^2 {{\partial r}\over{\partial s}} \e{ldensity}
\ee
\be
P(r,t) = \rho(r,t)kT/m \e{leos}
\ee
\be
{{\partial v}\over{\partial t}} = -{{1}\over{\rho(s,0)}}
\bigg({{r(s,t)}\over{s}}\bigg)^2 {{\partial (P+q)}\over{\partial s}} 
- {{GM(r)}\over{r}}^2  \e{lvelocity}
\ee
\be
{{\partial {\cal E}}\over{\partial t}} = 
{{P+q}\over{\rho}}{{\partial \rho}\over{\partial t}} - {{L}\over{\rho}} \e{lenergy}
\ee

In these equations, $s$ is the original position of each Lagrangian cell,
$r(s,t)$ is the corresponding position at time $t$,
$M(r)$ is the mass contained within radius r, 
${\cal E}$ is the internal energy of the gas,
${L}$ is the radiative gas cooling rate defined in equation \ref{eq:gasequil},
and $q$ is the Richtmyer-von Neumann pseudo-viscosity,
\be
q = \ell^2 \rho \bigg({{\partial v(s,t)}\over{\partial s}}\bigg)^2 {\ \rm when\ }
{{\partial v}\over{\partial s}} < 0
\ee
\be
q = 0{\ \rm when\ }
{{\partial v}\over{\partial s}} > 0
\ee
with $\ell$ a constant that is approximately the shock smoothing width over the grid spacing. 
These equations are
discretized as in \cite{Richtmyer1957}. 

In our calculations of the hydrodynamic evolution of the cores, we start
with a core 
in hydrostatic equilibrium and in radiative equilibrium.
In the subsequent hydrodynamic evolution, the temperature of the gas is calculated
at each time step taking into account the PdV work of compression or
expansion and the rate for radiative cooling, ${L}$ (equation \ref{eq:gasequil}).
Using the parameterized approximations for the gas cooling,
the change in internal energy may be calculated for each grid
cell using only the local values of density and velocity (compression) of the gas, 
and the temperatures of the gas and dust. The dust temperature itself depends
on the global structure of the core -- the angle averaged optical depth to 
the surface from each radial point. However, because the larger scale structure
of the core changes relatively slowly compared to the time step of the
hydrodynamic evolution, set by the Courant condition, the angle averaging
to determine the dust temperature need not be computed at every time step.
In addition, because of the approximation that the dust temperature is set
only by the radiative equilibrium of the dust and is independent of the
gas temperature, the dust temperature must be a smoothly varying function
in the core. Thus the dust temperature can be computed less frequently and
on a coarser grid and interpolated onto the hydrodynamic grid.

The initial core always contains some perturbations that cause some
initial movement of the gas in the core. These pertubations, at a fraction
of a percent level, are primarily due to coarse gridding
of the dust equilibrium and are on the scale of the coarse grid, about
1/20 the size of the core. The perturbations arise because
the initial dust and gas temperatures,
exactly match conditions of hydrostatic and radiative equilibrium
on the coarse grid,  but the interpolated dust temperatures and thus 
the interpolated gas temperatures do not exactly match the gas
temperature required for hydrostatic equilibrium between grid points. 
In both stable and unstable clouds the 
perturbations are smoothed in a time scale, $\delta/c_s \sim 10^4$ yrs
by small changes in temperature and density that quickly establish
radiative and hydrostatic equilibrium. 
In the gravitationally stable clouds,  once the perturbations are smoothed
out, no further evolution takes place.
In gravitationally unstable cores, 
even though the initial pertubations are quickly smoothed, the core is 
unable to establish an exact equilibrium and after a long time, 
several sound crossing times or tens of free fall times, the core
will begin gravitational collapse. The length of time for the core to
begin collapse is simply the time required for very small perturbations
to reach any appreciable magnitude. Once the inward velocities reach
even a few hundreths of the sound speed,
the core collapses in about a free fall time, 
$t_{{f}{f}} \sim 1/\sqrt{G\rho} \sim 10^5$ yrs.  Were it not for these
perturbations deriving from the coarse gridding of the radiative transfer,
it would be necessary to introduce perturbations into the initial structure
so that the unstable cores would collapse before patience with the
hydrodynamic evolution was exhausted.

\section{Radiative Transfer and Chemistry}

After the numerical hydrodynamic evolution of the cloud has been completed, the
temperature, density, and velocity as a function of time
are transferred to
a non-LTE accelerated $\Lambda$
iteration (ALI) radiative transfer code to determine model spectral line profiles 
for comparison with observations
(\cite{Keto2004}). 
The model cores are assumed to be at 150 pc,
the distance to the Taurus star-forming region and to be
observed with a telescope with a beam width of 20$^{\prime\prime}$.
Because the core radii 
are on the order of 10 beams, the resolution
is not critical except in the cases of the unstable cores which
have rapidly increasing density and velocity gradients in their centers.

To compute the spectral line profiles we also need to know
the abundances of the molecules in addition to the information provided 
by the hydrodynamical
evolution.  
The gas phase abundances of molecules are variable within the core because 
in the cold dense interiors of the dark cloud cores,
molecules freeze out of the gas phase onto the surfaces of dust grains 
at different characteristic densities according to their different volatilities
\citep{BrownCharnleyMillar1988, WillacyWilliams1993, HasegawaHerbstLeung1992, HasegawaHerbst1993, Caselli1999, Caselli2002a, Caselli2002b, Bergin2002, Tafalla2002, BerginLangerGoldsmith1995, Aikawa2001, Caselli2002c}.  
We follow \cite{Bergin2001}, \cite{Tafalla2002}, and \cite{Tafalla2004} and approximate 
the dependence of the abundance
on the density, $n$ as an exponential. 
\be
X(n) = X_0 \exp (-n/n_d)
\ee
with parameters for CS of abundance $X_0 = 10^{-8}$ and density $n_d = 10^{4}$ cm$^{-3}$. 
A more complete treatment of the molecular abundances of the starless cores derived from detailed
chemical modeling is discussed in Lee, Bergin \& Evans (2004). 
That study concludes that a
simple parameterized model of depletion, as adopted in our simulations, is a reasonable
approximation to their more detailed analysis.

\section{Model Cores}

We adopt as a standard model a core with a total mass of 5 \msun\   
and consider its evolution under a variety of conditions that
could be expected in and around the dark cloud cores. 
For a given mass, the central density determines
the gravitational stability of a non-isothermal BE sphere.
We model the evolution of cores with a range of central densities 
$n_{\rm H_2} = 10^3 - 10^7$ cm$^{-3}$ centered around the critical density for gravitational
instability, approximately $n_{\rm H_2} \sim 10^4$ cm$^{-3}$ for a core of 5 \msun . 

A core could be subject to a variation in external pressure 
from a nearby supernova, bipolar outflow, or the
turbulence of the interstellar medium. 
A change in external pressure 
could cause oscillations in a core, and a sufficient increase in external pressure
could cause a previously stable core to collapse.
Accordingly, we model the evolution of some cores in response to
changing external pressure.

The dark cloud cores may have some component of magnetic
or turbulent support in addition to thermal support. 
In a one-dimensional model, this additional support
can be approximated by modifying
the equation of state to include a non-thermal pressure.
This assumes that the non-thermal or magnetic energy has
a characteristic scale much smaller than the core so that
the effect is at least approximately the same as an isotropic
pressure. This assumption follows from estimates that show that the
large scale magnetic fields in cores are too weak
to affect their evolution (Nakano 1998). Thus we may 
assume that the non-thermal pressure is small scale and isotropic.
If the non-thermal pressure is initially proportional to density, then
the initial structure of the core may be determined as before with the 
Lane-Emden equation, but with a higher effective temperature than
in the case of support only through thermal pressure.
The subsequent evolution of the core depends on the equation of
state of the non-thermal pressure. For example, the cores that we
modeled with a non-thermal
adiabatic index greater than 4/3 behave as expected and 
never collapse to a point source, but form an
inner core of approximately constant density supported by non-thermal pressure.  As the
core evolves,
the surrounding envelope continues to collapse onto this inner core creating 
a steep density gradient and shock at the boundary of the inner core. 
Such a boundary region is never seen in observations of dark cloud
cores suggesting that the molecular gas in the dark cloud cores must have
a lower value of the adiabatic index.
In an analysis of the virial theorem, 
\cite{McKeeZweibel1995} suggested that the adiabatic index for magnetic
turbulence should be between 3/2 and 1/2 depending on the time and length
scales of the turbulence. A particularly convenient choice is an adiabatic
index of unity. This of course matches the adiabatic index of the thermal 
pressure, and the
dynamics are therefore similar to a thermally supported core with a higher temperature. 
However, the non-thermal pressure remains proportional to density only, while
the thermal pressure, being proportional to the temperature as well as the
density, will change as the temperature is changing.

We ran about 30 different numerical simulations to explore different combinations
and ranges of parameters.
Densities considered in our study include those:
\begin{enumerate}
\item expected to be stable
\item near the critical density of the non-isothermal BE sphere
\item expected to be unstable 
\end{enumerate}

Different equations of state considered in this study are those with:
\begin{enumerate}
\item thermal pressure only
\item thermal pressure with a non-thermal pressure initially proportional to
density
\begin{enumerate}
\item an adiabatic index of non-thermal pressure equal to unity
\item an adiabatic index of non-thermal pressure equal to 4/3 
\end{enumerate}
\end{enumerate}

Different external conditions applied to the cloud were:
\begin{enumerate}
\item constant external pressure
\item increasing external pressure
\begin{enumerate}
\item external pressure remaining below the critical pressure of the
a non-isothermal BE sphere
\item external pressure sufficiently high to cause collapse
\end{enumerate}
\end{enumerate}

Hydrodynamic models were computed for combinations of all the cases above. Despite the 
variety of initial conditions and equations of state, the behavior and properties of
the stable starless cores are quite similar to each other as are the behavior and properties
of the unstable pre-stellar cores.
In the sections below, we discuss in detail two models to illustrate the characteristics
of the two classes of cores.
We model the unstable pre-stellar cores with a hydrodynamic simulation that 
starts from an initial configuration in unstable equilibrium and evolves toward
free-fall gravitational collapse. The stable starless cores are modeled with
a simulation that starts in stable equilibrium and is perturbed into oscillatory 
contraction and expansion by a sudden increase in the external pressure.
Lastly, in order to illustrate
the thermal instability that is related to the critical density for gas-dust
collisional coupling we discuss the evolution of a core
with an initial configuration that is both gravitationally and thermally unstable. 
The initial expansion of this core caused by the thermal instability results in
a stable core undergoing damped oscillations of expansion and contraction.

\subsection{Hydrodynamic Evolution of an Unstable pre-Protostellar Core}

The theoretical evolution of a gravitationally unstable core is dynamically as 
expected from previous
simulations:  
small numerical perturbations grow for several crossing times (several Myr) until
they are sufficient
to tip the core off its initial unstable equilibrium toward collapse
\citep{Hunter1977, FosterChevalier1993}. 
Since the cooling time is short compared to the dynamical time, 
the temperature profile is set almost entirely by radiative equilibrium.  
Figures \ref{fig:unstableEvolPlot0} through \ref{fig:unstableEvolPlot39} 
show the properties of the core as the collapse
proceeds. Also shown are accompanying molecular line spectra as would be 
observed toward the center of the core.

Figures \ref{fig:unstableEvolPlot0} through \ref{fig:unstableEvolPlot39} 
show that the approximate force balance between pressure and gravity
that is set in the initial configuration
is maintained until the infall velocities begin to approach the sound speed.
Thus
the density structure continues to approximate that of a pressure supported
BE sphere as the core evolves toward free-fall collapse. 

Our hydrodynamic simulation,
as well as analytic considerations  \citep{WhitworthWT2001},
show that gravitational collapse starting from an
equilibrium configuration BE sphere is characterized by inwardly increasing velocities.
A velocity profile with velocities that increase inward is 
not unique to the density profile of unstable
BE spheres, but is still a diagnostic inasmuch as other initial configurations 
such as the uniform spheres considered in Jeans instability and the
well studied 
singular isothermal spheres \citep{Shu1977} and singular logotropic spheres
\citep{McLaughlinPudritz1997} evolve to collapse with very 
different velocity profiles.
Inwardly increasing velocities are indicated in
molecular line observations of the cores L1544 and L1521F
by the velocity gradient inferred from the details of the
observed spectral line shapes and from
the widths of spectral lines of volatile molecular species that increase
toward the centers of the cores.
Figures \ref{fig:unstableEvolPlot0} through \ref{fig:unstableEvolPlot39} 
show how the spectral profile of \diaz\
develops the characteristic asymmetric split with a stronger blue peak
as the infall velocity increases. 
This characteristic split is seen
in the observed
\diaz\ spectrum from L1544, for example in figure 2 of
\cite{Williams1999} and in figures 4 and 5 of \cite{Keto2004}. 
An increase in the linewidth of N$_2$D$^+$ is observed toward the centers of
the cores
L1544 and L1521F \citep{Caselli2002b,Crapsi2004,Crapsi2005}. Although
the observations of line width do not indicate the direction of the velocity, the
increase in line width is indicative of an increase in the magnitude of the
velocity toward the center of the core.

\subsection{Hydrodynamic Evolution of a Stable, Oscillating Starless Core}

Hydrodynamic simulations of cores in stable equilibrium show no evolution
unless the core is subjected to a perturbation. While observations
of many cores do not show velocities above the observational detection limit,
about one tenth the sound speed, other cores indicate expanding
or contracting motions or more complex combinations indicative of
non-spherical perturbations. 
While some of these cores may be unstable and evolving toward collapse as described
in the previous section, others may be oscillating around an equilibrium mean.
The non-spherical morphologies of
many cores, often showing aspect ratios of 2:1, may indicate 
oscillations about an equilibrium mean which may be stable or unstable.

In our example model of a stable starless core we begin the hydrodynamic
simulation in hydrostatic and radiative equilibrium.
The core starts with central density of
$2\times 10^3$ cm$^{-3}$, well to the stable side of the Bonnor
criteria for non-isothermal spheres so that an
increase in external pressure by a factor of 1.5, held constant thereafter,
pushes the core close to but
not past the maximum pressure for 
stability. 
The increase in pressure
propagates into the core, eventually reflecting off the center, while
the core as a whole adjusts its density profile to a new equilibrium with
the increased external pressure. An initial overshoot of the new
equilibrium results in damped
oscillations of contraction and expansion.
The oscillations imply that inertial forces
are important in the dynamics, but the inertial forces are still small
compared to the pressure and gravitational forces so that the core
remains in approximate force balance throughout the evolution. 

Figure \ref{fig:stableEvolPlot0} shows the structure of the core and the accompanying
spectra when the
head of the velocity and pressure wave has advanced half way
into the cloud at elapsed time 1.003 Myr. 
Because of the low density,
the gas cools efficiently throughout the volume of the
cloud by molecular line radiation and
the gas temperature is nearly constant. 
The predicted spectral line profiles of an oscillating core are particularly
interesting in comparison with observed profiles.
The CS(1-0) spectrum shows an asymmetric
profile with a stronger blue peak that is characteristic of inward motion.
In figure \ref{fig:stableEvolPlot8} at 1.613 Myr, the wave has 
reflected off the core and is headed outward. At this time, most of the
core is still moving inward, responding to the increased external pressure, and
the spectral line profile of CS still shows an asymmetric line split 
characteristic of inward motion. By a time
of 3.705 Myr (figure \ref{fig:stableEvolPlot28}), when the contraction 
of the core has overshot the equilibrium point, 
the whole core is expanding outward. The spectral line profiles of
CS now show an asymmetric split line with a stronger red peak that is
characteristic of outward motion. 
Observations of CS spectra in B68 show asymmetric split spectra of both
senses, inward motions and outward motions, at different positions in
the cloud, for example, figure 6 of \cite{Lada2003}. 

Model spectra of \diaz\ from the same hydrodynamic simulation show only
symmetric profiles with no asymmetric splitting. The difference between
the CS spectra and the \diaz\ spectra relates to the differences in the
velocity fields in the regions where each of the molecular lines is
generated. In the dense centers of the starless cores the CS molecule 
is depleted from the gas phase by freeze-out onto dust grains, but the
more volatile \diaz\ molecule maintains a constant abundance. Because the line
brightness for any collisionally excited molecule scales with the square
of the density, the \diaz\ line is formed predominantly in the dense 
gas in the center of the core whereas the CS line is formed predominantly
in the envelope. In the breathing mode oscillations such as those induced
in our simulation by an increase in the external pressure, the highest 
velocities are generally found in the envelope while the velocity in the
core remains low. Thus the theoretical model indicates that the CS molecule
will show an asymmetric split spectrum while the \diaz\ molecule will
not. 

The theoretically predicted line shapes of the volatile and depleting species
are observed in L1517B.
Observations of \diaz\ in L1517B, figure 13 of \cite{Keto2004} and
figure 6 of \cite{Tafalla2004} show only narrow symmetric spectra
while the spectrum of HCN (another species strongly depleted in dense gas) 
shows a pronounced asymmetric split with a higher
blue peak indicating expansion \citep{Sohn2004}. 
Based on the modeling, L1517B would
be in the expanding phase of an oscillation.
Observations of B68 in \diaz\ and CS also match the theoretical prediction with
a nearly symmetric profile observed in the volatile \diaz\ molecule and significant
splitting observed in the more refractory tracer, for example figure 2 of  \cite{Lada2003} 
and in \cite{Redman2005}.

In general, the stable starless cores show
warmer temperatures, lower densities, less central concentration of their density
profiles, and a variety of spectral line shapes including both expansion and
contraction. 
Since the velocities in the centers of the stable starless cores are very
low, the spectra of volatile tracers show no increase in line width at 
positions closer to the core center. In contrast, the increased
line widths of volatile tracers observed toward the centers of 
unstable, pre-stellar cores have much higher velocities in their centers
than in their envelopes.
This combination of properties is sufficient to distinguish these cores 
that are not currently evolving to form protostars 
from those in the unstable category that are in the earliest stages of protostellar
formation.

\subsection{A Thermally Unstable Core}

Under the right conditions,
a core may be subject to a thermal instability if as the core expands,
the collisional coupling that allows the dust to cool the gas is
reduced causing the gas to warm and further drive the expansion.
Since the gravitational stability is determined by the total mass of
the core as well as the density while the gas-dust collisional
coupling is a function of the density alone, 
a core may be unstable to gravitational collapse 
but thermally unstable to expansion.
As a last example we follow the evolution of a model core 
whose initial configuration is doubly unstable.

The central density in the initial
equilibrium configuration is $10^5$ cm$^{-3}$ placing it on the gravitationally unstable
portion of the Bonner stability curve in figure \ref{fig:pvplot_nonthermal_pressure}. 
However, because the
density is very close to the critical density for the collisional coupling of
the gas and dust, the core is thermally unstable to expansion. As the core
expands, the cooling of the gas is reduced as the transfer of energy from the gas
to the dust through collisions becomes less effective. 
The gain in internal energy, $dQ/dt$, then
furthers the expansion. 
The increase in energy from the reduced cooling
is related to the gas temperature and the work done in expansion,
\be
{{dQ}\over{dt}} =
{{d}\over{dt}}\bigg({{3kT}\over{2m}}\bigg)  + P {{dV}\over{dt}} 
\e{internalenergy}
\ee
The numerical simulation (figure \ref{fig:TDPThermalInstability})  
shows that as the cloud expands, the
gas temperature increases while the density (volume$^{-1}$) decreases.
This is seen in the first cycle of expansion, at times less than 2 Myr
and for densities less than $10^{4.5}$ cm$^{-3}$, when the temperature 
increases with decreasing density, the opposite of the behavior expected
for a gas with an adiabatic equation of state. After this time, the density
is low enough that the gas is no longer collisionally coupled to the
dust, and the gas behaves approximately adiabatically with the
temperature and density increasing or decreasing together. 

The core resolves its doubly unstable state by expansion to a lower 
density state that is both gravitationally and thermally stable.
Expansion is favored over collapse in this model core even though
the linear analysis of \cite{Field1965} predicts
that both expansion and contraction are equally unstable as long as the
perturbations are small.
Although the linear thermal instability itself is symmetric, in the model core the
asymmetry toward expansion derives from non-linear effects associated with
the radiative equilibrium of the dust. For example, in the initial state of the
model, the gas is collisionally well coupled to the dust and the temperatures
of the gas and the dust are nearly the same. An increase in density,
does not cause the gas to become cooler because increased collisional coupling
cannot reduce the gas temperature below the dust temperature. The dust 
temperature remains approximately constant because density changes localized in the
center of core do not significantly affect the overall radiative 
equilibrium and temperature.  On the other hand,
a decrease in the gas density in the center of the core 
causes an immediate rise in the local gas temperature 
because of the reduced collisional coupling. Thus the instability appears
one-sided.

The thermal instability is also evident in the 
variation of pressure with temperature and density. For an adiabatic
change, the three quantities are related as $P = k\rho^{\gamma} $  and
$T = k\rho^{\gamma -1}$ where $\gamma$ is the ratio of
specific heats and $k$ is a constant.
A plot of the log of the temperature 
as function of the log of the density (figure \ref{fig:effAdiabaticIndex}) shows
that at low density the slope of the curve is negative. 
If the relationship between $T$ and $\rho$ were expressed in the
form of the adiabatic relationship, then $\gamma^\prime < 1$ where $\gamma^\prime$ is the exponent
derived from the empirical relationship between $T$ and $\rho$ rather than 
the ratio of specific heats
as in the
adiabatic relationship.
However,
on the portion of the log $T$ versus log $P$ curve 
corresponding to the initial stage of the expansion when the gas is at higher density 
the slope of the curve is positive and $\gamma^{\prime}$ must be greater than unity.
A corresponding plot of the gas pressure versus density
would show that 
$\gamma^{\prime\prime}$ 
in the relationship $P \sim \rho^{\gamma^{\prime\prime}}$,
is positive and nearly constant at all
times in the evolution. 
This is because the relationship between pressure and density is set primarily by the equation of state
of the gas, equation \ref{eq:leos},
whereas
the relationship between temperature and density 
is controlled primarily by
radiative equilibrium.

This simulation of thermal instability demonstrates how 
the relationships between pressure, temperature, and density,
can change dramatically at densities that are
approximately the same as the critical densities for gravitational instability
in cores of a few solar masses. 
The ability of the gas to change its effective equation of state may be important
in facilitating the formation of gravitationally stable cores 
\citep{BPVSScalo1999, Jappsen2005}. 
The thermal instability improves the chances for
the formation of
gravitationally bound and yet stable cores since the instability prevents
gravitational collapse of some cores that would otherwise be unstable.
For example, cores of 5 $M_\odot$ with initial central densities between $10^5$ and $10^6$ are unstable
to thermal expansion and do not collapse. 
Cores with 5 M$_\odot$ and
lower densities are gravitationally stable  as expected. 
Cores with 5 M$_\odot$ and
densities higher
than $10^6$ always collapse.

\section{Conclusions}

A comparison of the observed and theoretically predicted properties of dark cloud cores
suggests that the cores are
gravitationally bound dynamical units with a density structure set by approximate force balance
between gravity and thermal pressure and a temperature structure set by radiative
equilibrium with the general interstellar radiation field.  
Such gravitationally bound structures might exist in hydrodynamic
equilibrium below the inertial range 
at the lower end of the supersonic turbulent cascade.  

The interstellar turbulence
should generate cores in a range of densities and masses, all below the sonic limit of
the turbulent cascade, some that are gravitationally
stable and some that are unstable. 
The stable cores are in general less dense, more
extended, and warmer. 
The unstable cores are denser, more centrally concentrated, and colder.
Their velocities are generally inward, progressing toward free-fall collapse.
Most of the observed cores 
can be classified into stable and
unstable categories on the basis of these properties.

Our hydrodynamic study shows
that cores may oscillate with periods
of the sound crossing time, about one million years, if subjected 
to a modest external perturbation of pressure.  
Thus stable starless cores could be long-lived in a
modestly turbulent environment. While the one-dimensional model cores are
permitted only radially symmetric oscillations, the dark cloud cores 
in a turbulent interstellar medium
would be expected to have higher order oscillations. 
For example,
the elliptical morphologies of many cores may be an expression of
slow, asymmetric oscillations about a mean spherical equilibrium.
Higher order oscillations have been inferred from observations of
B68 \citep{AlvesLadaLada2001}.

There is no evolutionary path from the stable to the unstable cores, thus not all
observed cores are evolving toward gravitational collapse and star formation. Stable
cores may persist indefinitely until perturbed by changing external conditions.

Spectral line observations of molecules with different degrees of volatility can be
used to distinguish the differences in properties in the centers of the cores and their
envelopes. 
The rate at which molecules freeze out of the gas phase onto dust grains is dependent on
the collisions with dust grains and hence the density. More volatile species remain in
the gas phase at higher densities. Spectral lines of these molecules are generated in,
and provide information on, the denser centers of the cores.
Less volatile species are severely depleted in the dense core centers, and the
spectral lines of these species are generated in and provide information on 
the envelopes of the cores.

Because the gas temperature is determined predominantly by radiative equilibrium,
the effective adiabatic index,  can change value and sign
as the density and optical depth of the cores change with expansion or contraction.
In particular, the rate of radiative cooling and hence the temperature of the gas 
are dependent on whether the gas density is above or below the critical density 
for effective collisional coupling of the gas and the dust. 
The reduction in the gas cooling rate at lower gas densities can create a thermal
instability that causes the gas to heat up as it expands further driving the
expansion.

\clearpage

\begin{figure}[t] 
\includegraphics[width=5in]{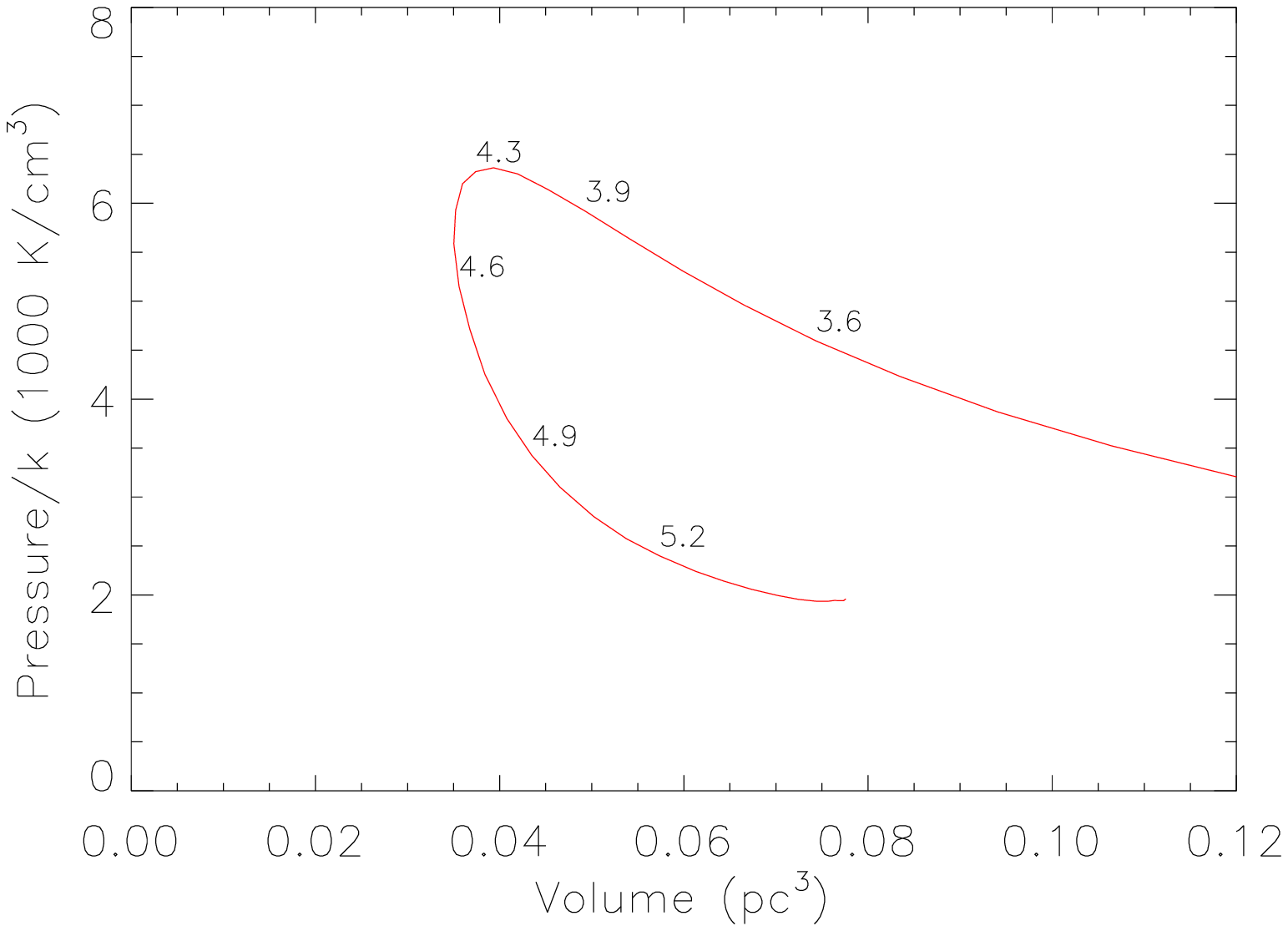}
\caption{ 
Curve of volume and boundary pressure of a 5 M$_\odot$ sphere in
hydrostatic equilibrium and radiative equilibrium parameterized
as a function of the gas density in the center of the
sphere. The log of the gas density is marked at points along the curve.
The spheres are supported against gravity by the thermal pressure
of temperatures
ranging from 7 to 14 K. The spheres with central densities, volumes, and
boundary pressures that are to the right of the maximum pressure
are in stable equilibrium. Those spheres on the curve 
to the left of the pressure maximum are in unstable equilibrium. 
}
\label{fig:pvplot_variable_T}   
\end{figure}
\clearpage

\begin{figure}[t] 
\includegraphics[width=5in]{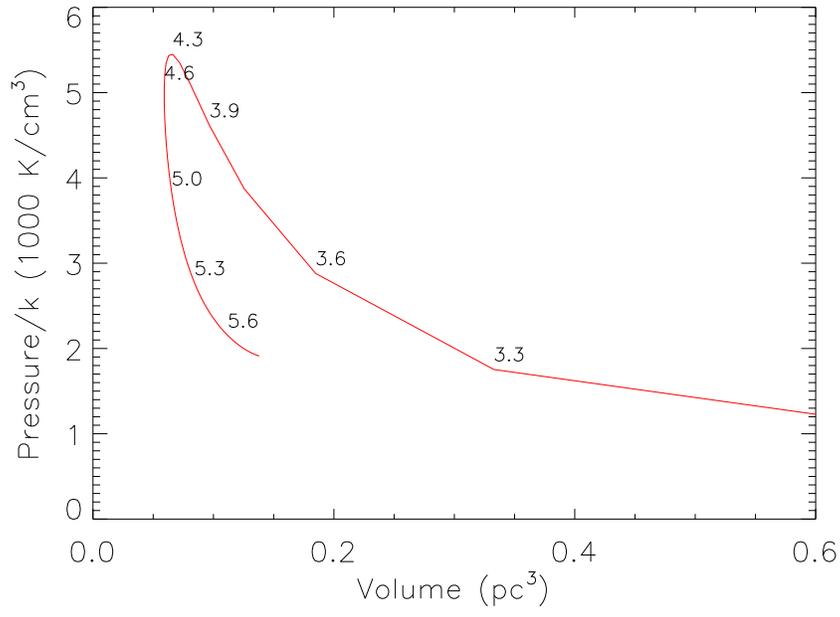}
\caption{
Same as \ref{fig:pvplot_variable_T}
except that the sphere is isothermal at 10 K.
}
\label{fig:pvplot_constant_T}   
\end{figure}
\clearpage
   
\begin{figure}[t] 
\includegraphics[width=5in]{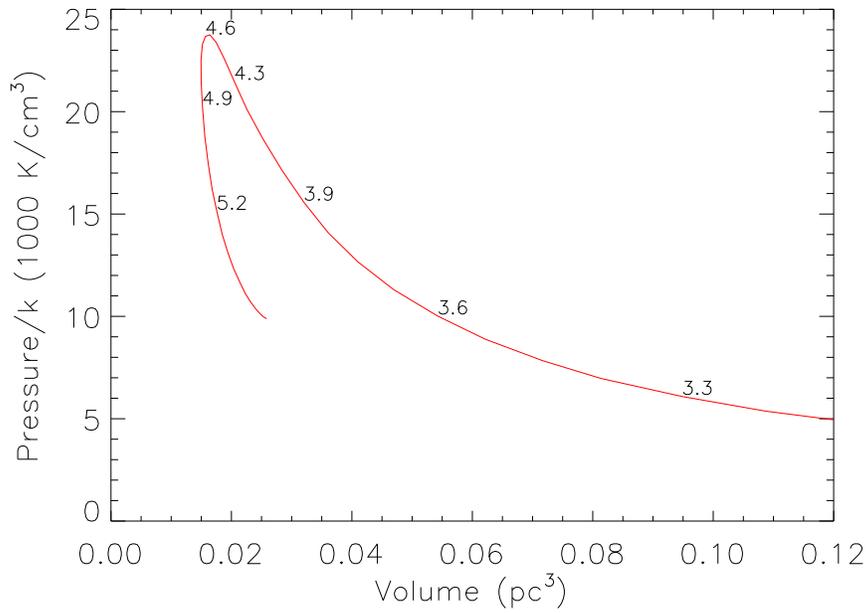}
\caption{Same as \ref{fig:pvplot_variable_T}, except that the gas has additional 
internal energy equal to one-quarter of the thermal energy. This
additional energy could come from small scale turbulence or magnetic
fields. With this additional internal pressure, the clouds are able to support
a higher maximum central gas density and boundary pressure.
}
\label{fig:pvplot_nonthermal_pressure}   
\end{figure}
\clearpage
   
\begin{figure}[t] 
\includegraphics[width=5in]{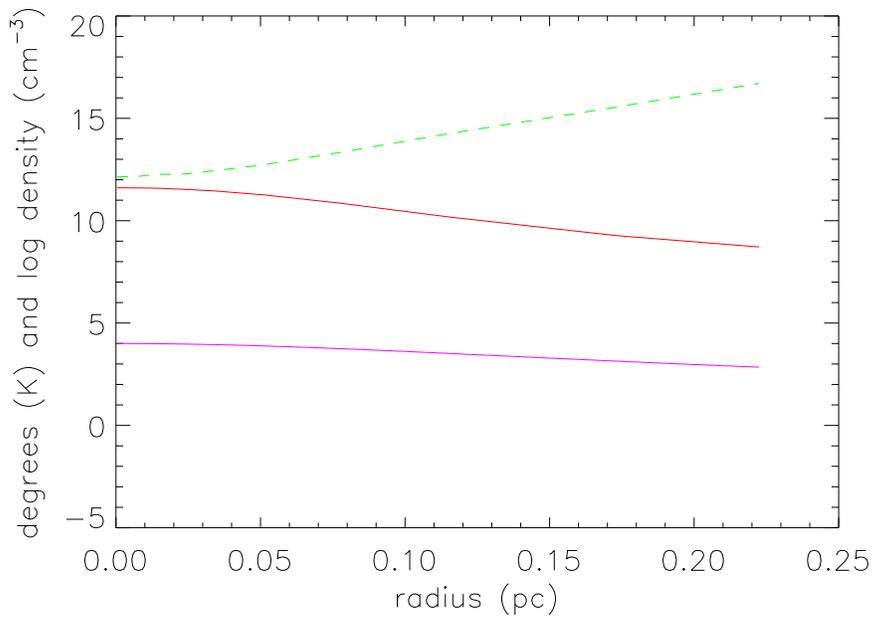}
\caption{Dust and gas temperature and gas density as a function of radius for
a sphere in radiative and hydrostatic equilibrium with a central density
of $10^4$ cm$^{-3}$. The log of the gas density 
is the lower line.  The two upper lines are the dust temperature (dashed line) 
and the gas temperature. The ordinate axis may be read as either the log of the
density in cm$^{-3}$ or the temperature in degrees K. The sphere is in stable
gravitational equilibrium.
}
\label{fig:tdv_stable}   
\end{figure}
\clearpage

\begin{figure}[t] 
\includegraphics[width=5in]{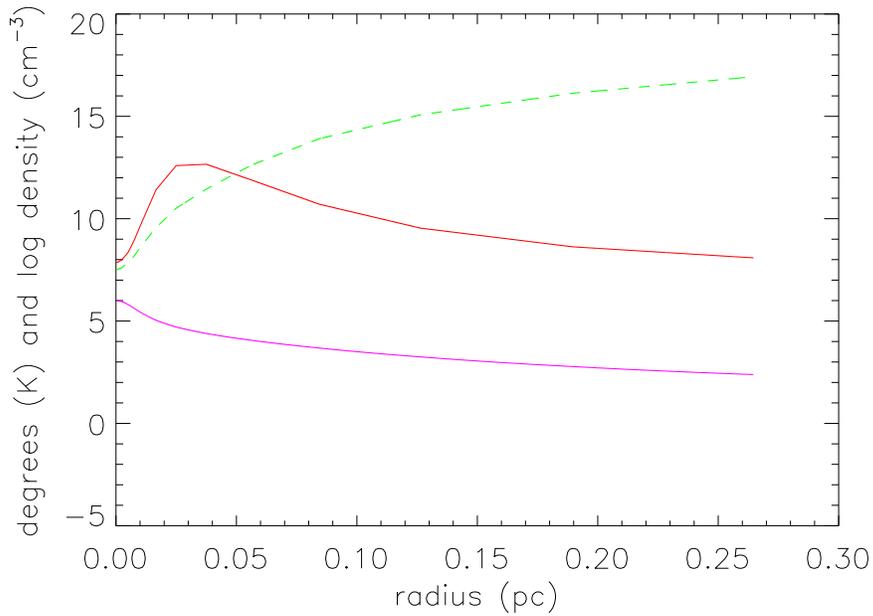}
\caption{Same as figure \ref{fig:tdv_stable} except that the central
density of the sphere is $10^6$ cm$^{-3}$. In this sphere the gas
temperature initially increases inward as the molecular line radiation that
cools the gas becomes ineffective because of the increasing optical
depth. As the gas density increases to the critical density for gas
dust coupling, the gas temperature approaches the dust temperature.
At the high gas density of this sphere, 
the sphere is unstable to gravitational collapse.
}
\label{fig:tdv_unstable}   
\end{figure}
\clearpage

\begin{figure}[t] 
\includegraphics[width=7in]{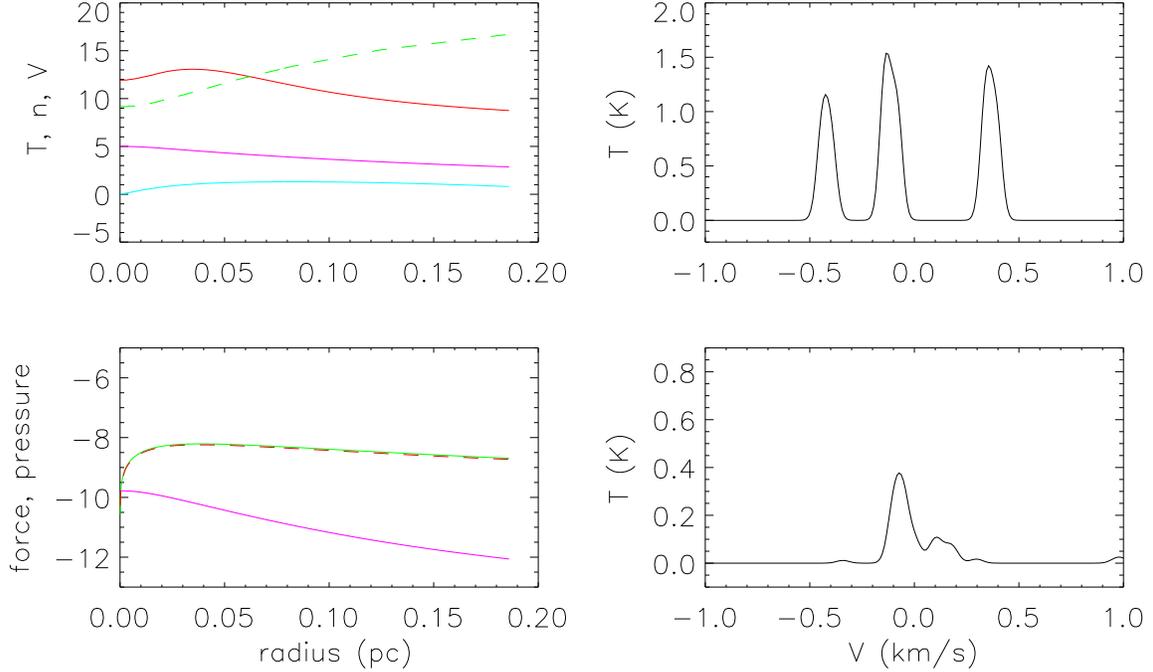}
\caption{
Figures \ref{fig:unstableEvolPlot0} through \ref{fig:unstableEvolPlot39} show
the evolution of a core starting from an initial configuration that is in
equilibrium, but unstable to gravitational collapse. In this simulation, the
cloud remains static for about 3 Myr, about 3 sound crossing times while
small perturbations originating from numerical noise build up to move the
cloud off its unstable equilibrium. Once the cloud begins to contract, the
velocities in the center of the cloud become supersonic in a few $10^5$ yrs
and the core is then in unsupported free-fall.\hfill\break
~
The upper left panel shows the density (log cm$^{-3}$), and gas and dust temperatures (K), in
the same format as figures \ref{fig:tdv_stable} and \ref{fig:tdv_unstable}, and  also the
velocity (sound speed times -10) as a function of the radius (log pc). 
For example, a velocity of 10 on the left
axis indicates an inward velocity equal to the sound speed. The lower left
panel shows the gravitational force (solid green upper line), the pressure force (dashed red upper line), 
and the pressure (lower solid pink line), in cgs units.
This panel (in the subsequent figures) shows that the cloud remains in approximate force
balance until the core begins free-fall collapse.\hfill\break
~
The two right panels show the spectral lines of \diaz (1-0) and \diaz (3-2) as they
would be observed in the model core. The radiative transfer simulation assumes the
core is at a distance of 150 pc and that the telescope beam is 20$^{\prime\prime}$.
Each panel shows the hyperfine lines that are within $\pm 1$ kms$^{-1}$. These are
the inner three hyperfines in the case of \diaz (1-0).
}
\label{fig:unstableEvolPlot0}   
\end{figure}
\clearpage

\begin{figure}[t] 
\includegraphics[width=7in]{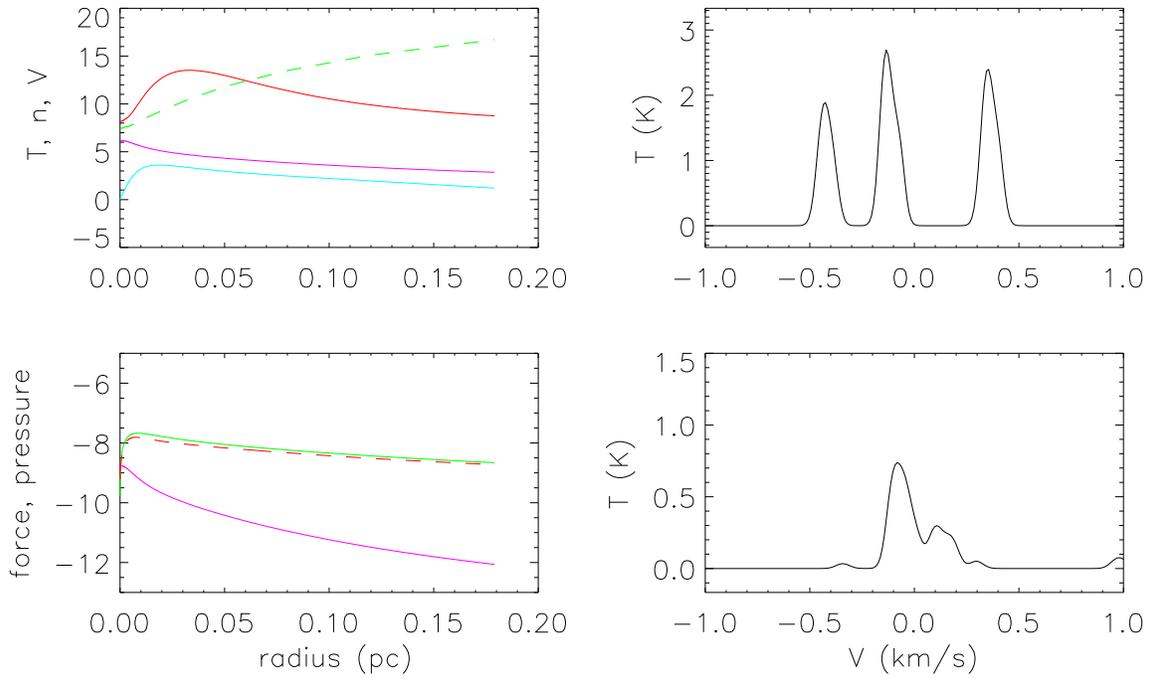}
\caption{
Same as \ref {fig:unstableEvolPlot0} but at an evolutionary time of 3.22 Myr.
As the gas density in the center increases, the gas becomes collisionally coupled
to the dust and the gas temperature approaches the dust temperature.
}
\label{fig:unstableEvolPlot30}   
\end{figure}
\clearpage

\begin{figure}[t] 
\includegraphics[width=7in]{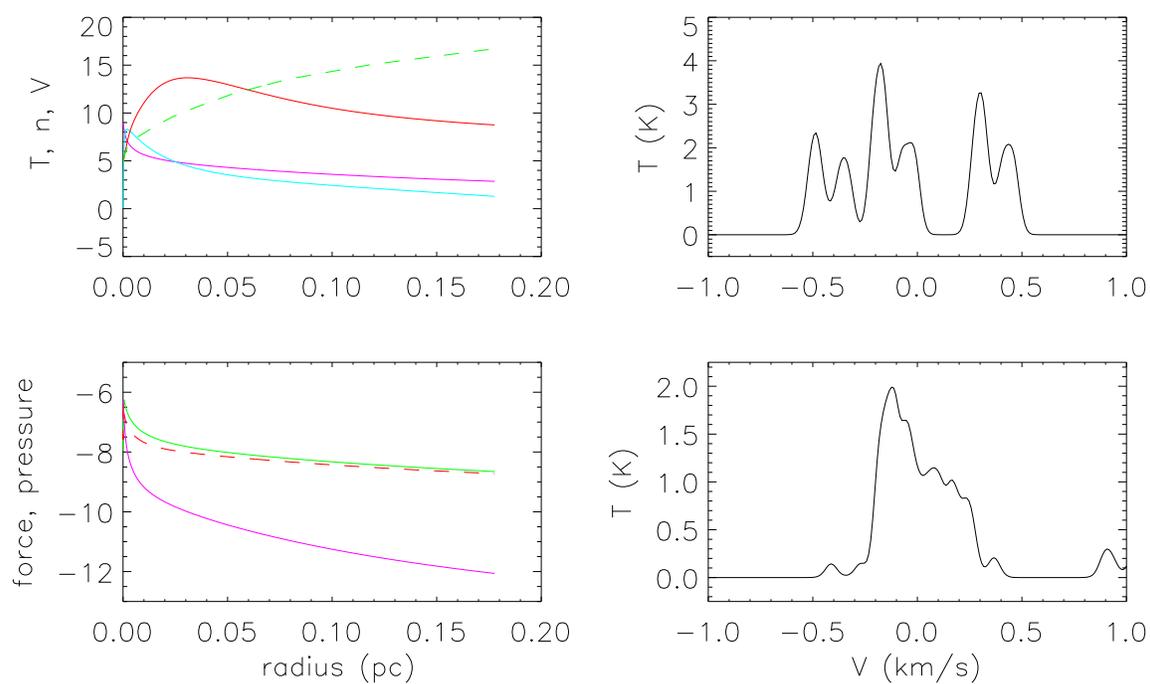}
\caption{
Same as \ref {fig:unstableEvolPlot0} but at an evolutionary time of 3.37 Myr.
As the velocity gradient steepens with the inward velocity approaching 
the sound speed in the center of the cloud, and the velocities still near zero
at the edge, the low velocity gas at the edge of the cloud absorbs the emission
from the center of the cloud creating a split spectrum. The profile of the
(3-2) line is dominated by numerous hyperfine components.
}
\label{fig:unstableEvolPlot39}   
\end{figure}
\clearpage

\begin{figure}[t] 
\includegraphics[width=7in]{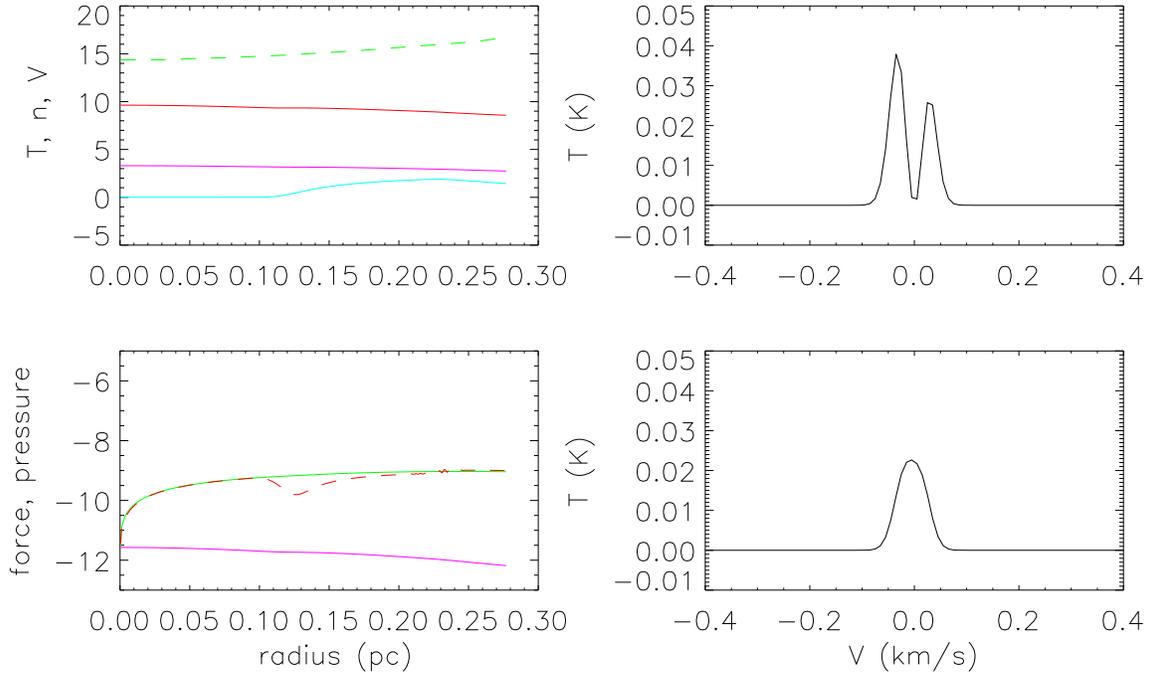}
\caption{
Figures \ref{fig:stableEvolPlot0} through \ref{fig:stableEvolPlot28} show
the evolution of a core starting from an initial configuration that is in
stable equilibrium, but subject to an increase in the external pressure
by a factor of 1.5. In this simulation, the core contracts to a new
equilibrium configuration in response to the increased pressure
through damped oscillations of contraction and expansion. 
This plot shows the model at an evolutionary time of 1.003 Myr
when the pressure wave has moved about half way into
the core. 
The four panels are as explained in figure \ref{fig:unstableEvolPlot0},
except that the spectral lines illustrated in this figure are CS(1-0) and
CS(3-2).
The asymmetric and split CS(1-0) spectrum with the brighter blue peak is
indicative of contraction. The (3-2) line does not show this splitting
because the line is optically thin.
}
\label{fig:stableEvolPlot0}   
\end{figure}
\clearpage

\begin{figure}[t] 
\includegraphics[width=7in]{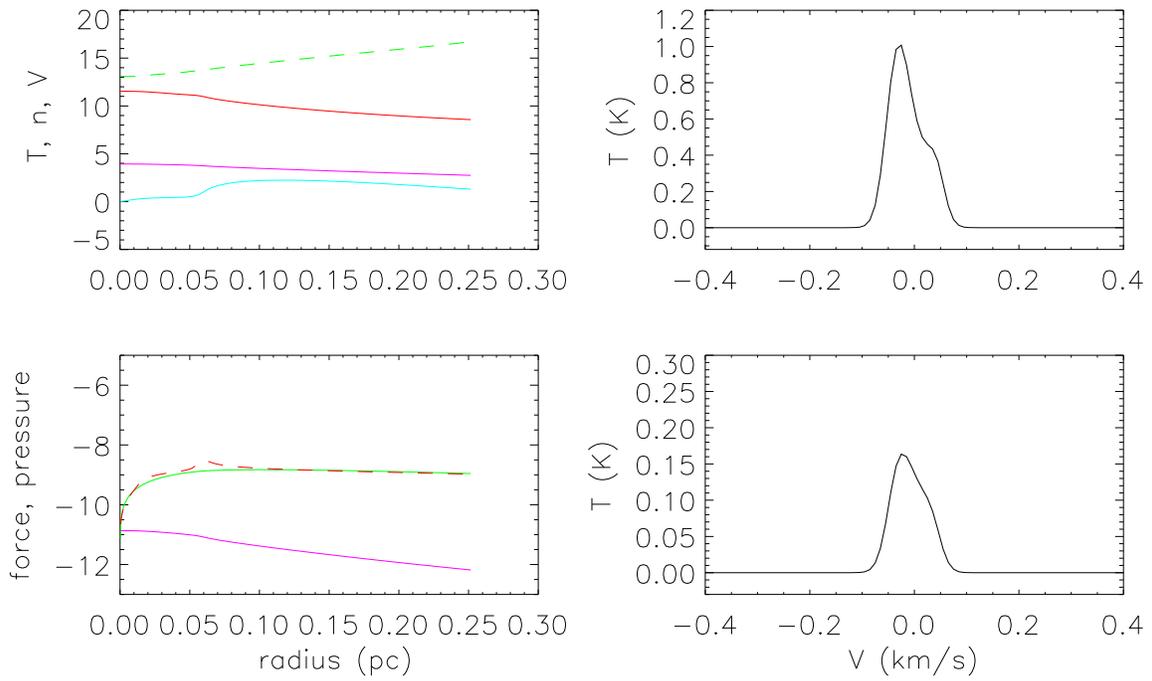}
\caption{
Same as \ref {fig:stableEvolPlot0} but at an evolutionary time of 1.613 Myr
when the pressure wave has reflected off the center of
the core. Even though the pressure wave is moving outward, most of the cloud
is still contracting in response to increased boundary pressure,
and the asymmetric CS spectra show a brighter blue peak 
indicative of contraction. The (3-2) spectrum shows asymmetry because at the
higher gas density, the line is becoming optically thick.
}
\label{fig:stableEvolPlot8}   
\end{figure}
\clearpage

\begin{figure}[t] 
\includegraphics[width=7in]{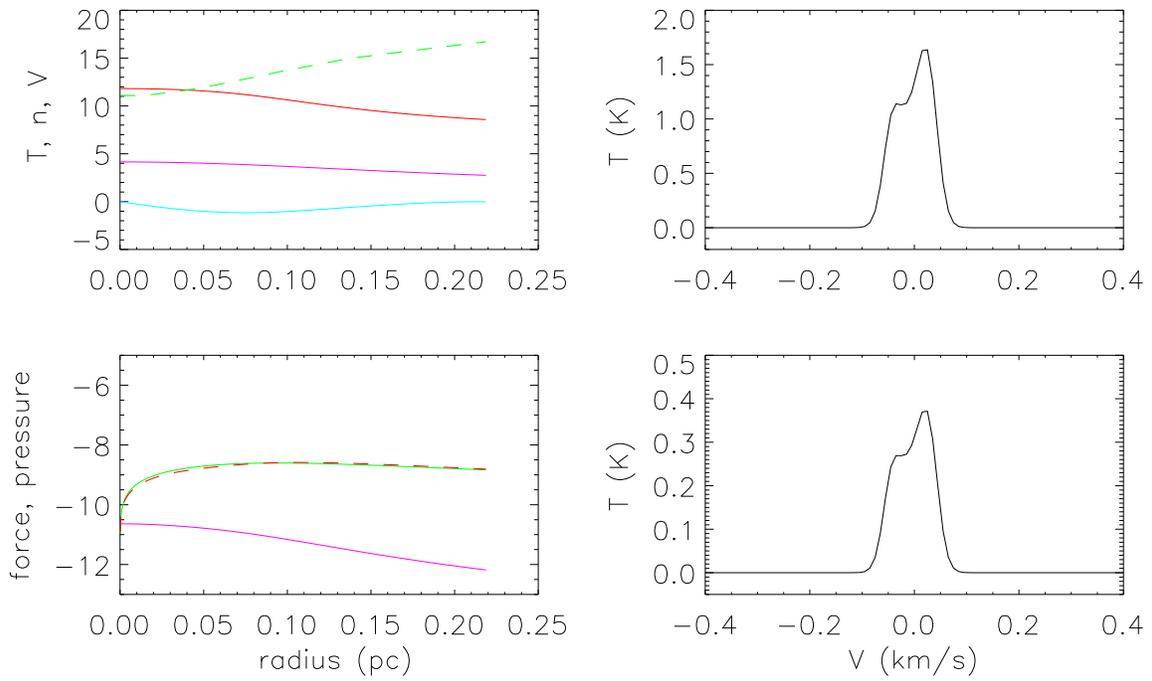}
\caption{
Same as \ref {fig:stableEvolPlot0} but at an evolutionary time of 3.705 Myr.
At this time most of the cloud is expanding having overshot the equilibrium
with the increased external pressure. The asymmetric CS profiles now show
a brighter red peak indicative of expansion.
}
\label{fig:stableEvolPlot28}   
\end{figure}
\clearpage

\begin{figure}[t] 
\includegraphics{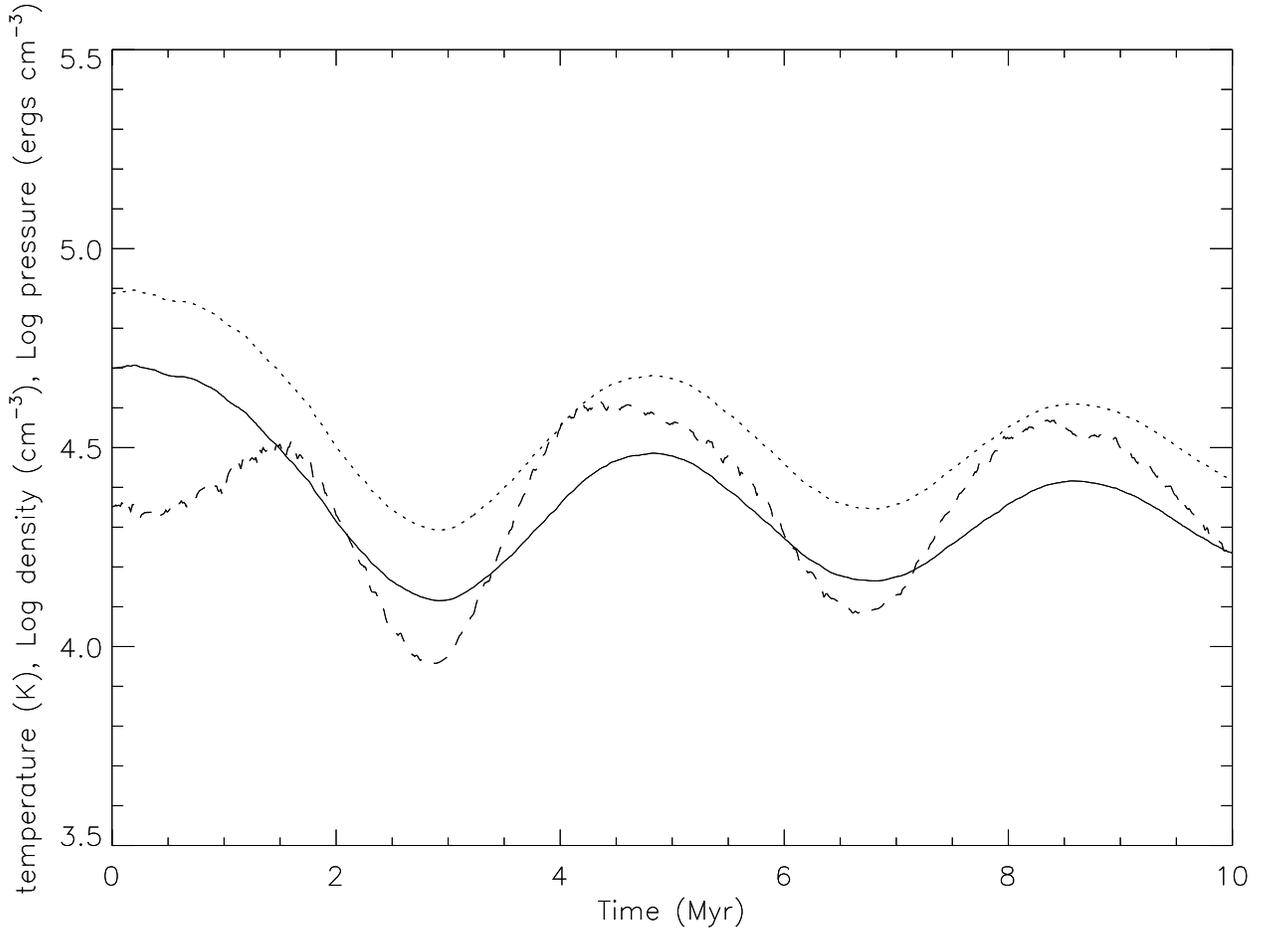}
\caption{
Temperature, density, and pressure as a function of time at the center of a core with an
initial configuration that is unstable to gravitational collapse and also unstable
to expansion by a thermal instability. The thermal instability is evident in the
early evolution of the cloud where the temperature can be seen to be increasing as
the density decreases. Once the cloud has expanded so that the density remains
below the critical density for gas dust coupling, the thermal instability no 
longer operates and the temperature behaves as for a gas with an approximately adiabatic
equation of state. \hfill\break
The log of the gas density (cm$^{-3}$) is the solid line, the log of the 
pressure (ergs cm$^{-3}$) is the dotted line, and the 
temperature is the dashed line. To align the curves vertically, 7 has been subtracted
from the temperature and 1 from the log pressure.
}
\label{fig:TDPThermalInstability}   
\end{figure}
\clearpage

\begin{figure}[t] 
\includegraphics{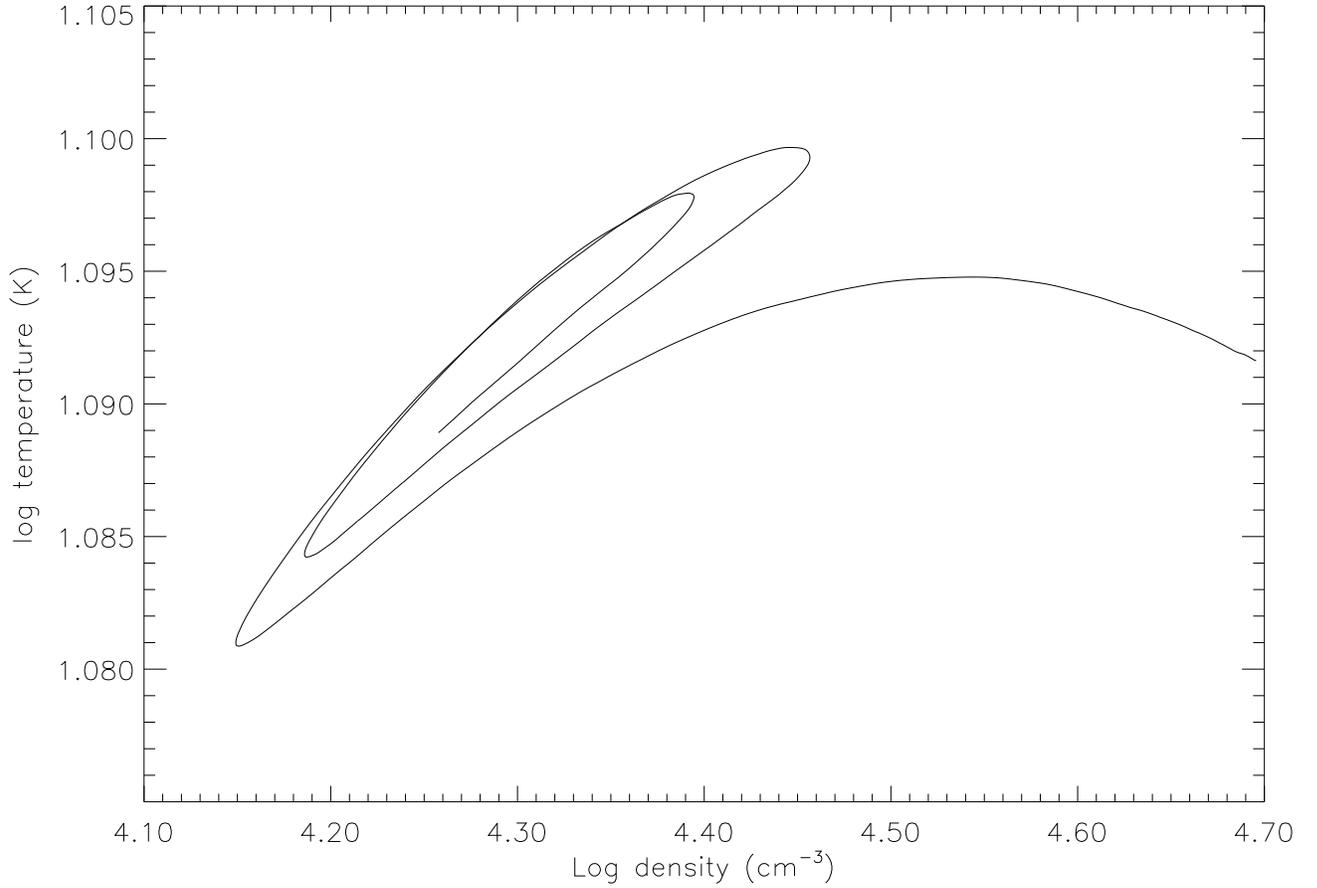}
\caption{
Temperature versus density at all times during the evolution of the thermally
unstable core wth the evolution illustrated in figure \ref{fig:TDPThermalInstability}.
The slope of the curve is the effective adiabatic index of the gas, $T \sim \rho^{\gamma-1}$.
The effective adiabatic index, given by the slope of the curve plus unity,  
evolves from negative to 
positive as the cloud expands.
}
\label{fig:effAdiabaticIndex}   
\end{figure}
\clearpage

\end{document}